# Quantum Information Processing with Ferroelectrically Coupled Quantum Dots

Jeremy Levy

*Department of Physics and Astronomy, University of Pittsburgh*

*3941 O'Hara St., Pittsburgh, PA  15260  USA*

*e-mail: jlevy@pitt.edu*

I describe a proposal to construct a quantum information processor using ferroelectrically coupled Ge/Si quantum dots.  The spin of single electrons form the fundamental qubits.  Small (<10 nm diameter) Ge quantum dots are optically excited to create spin polarized electrons in Si.  The static polarization of an epitaxial ferroelectric thin film confines electrons laterally in the semiconductor; spin interactions between nearest neighbor electrons are mediated by the nonlinear process of optical rectification.  Single qubit operations are achieved through "g-factor engineering" in the Ge/Si structures; spin-spin interactions occur through Heisenberg exchange, controlled by ferroelectric gates.  A method for reading out the final state, while required for quantum computing, is not described; electronic approaches involving single electron transistors may prove fruitful in satisfying this requirement.

PACS: 73.20.Dx,  77.84, 03.67.L

# I. Introduction

Quantum information processing is a central component for quantum computation[1], cryptography[2] and secure communication over large distances[3]. A quantum information processor must generate or accept an entangled quantum state as input, perform controlled operations on that state, and produce an output for measurement or further processing. The requirements for implementing such a system are stringent, and the technical challenges great. The ultimate form that a successfully engineered quantum information processor takes will determine, in large part, the architecture of technologies that utilize these new capabilities.

Ground breaking experiments by Kikkawa *et al.* have given hope that semiconductor-based architectures that utilize the spin of electrons will enable these requirements to be met. Experiments in direct band semiconductors such as GaAs have shown that electron spins exhibit a long-lived coherence that persists to room temperature[4] and is robust against diffusion[5] and transport across interfaces[6]. However, electronic spin lifetimes in III-V and II-VI systems are limited by spin-orbit and hyperfine coupling. Si is an attractive material for spin-based quantum computation[7] because of weak spin-orbit coupling and because Si is primarily composed of isotopes with nuclear spin zero; consequently decoherence times of the electron spin can be extremely long.

A number of solid state spin-based approaches to quantum information processing have been proposed in which localized states are manipulated through externally applied gate voltages[7-9]. Still, several hurdles remain. The control gates, especially those that mediate interactions between electron spins, must be modulated extremely rapidly and accurately, near or beyond the limit of what is achievable with current technology. Because the separation between bound states must be comparable to the Bohr radius for the electron, extremely high-resolution lithography is also required in most cases.

Here I describe an approach to quantum information processing that uses the static and dynamic polarization of a ferroelectric thin film to control electron spin interactions in Si. Two properties of ferroelectric materials are involved: (1) the use of static ferroelectric domains that confine laterally electrons in an adjacent semiconductor, and (2) a large nonlinear response (optical rectification) that enables terahertz polarizations to develop and influence nearby electronic states.

## II. Requirements for Quantum Computing

The requirements for quantum computing are exceedingly stringent, and have been delineated by a number of authors[1,10,11]. Here I follow the exposition of DiVincenzo[10], who outlines five requirements, summarized in the first column of Table I. The second column of Table I summarizes how these requirements will be met using ferroelectric/semiconductor heterostructures. The remainder of this paper will expand on the proposal as outlined in Table I.

**Requirement (R1)** states that there must be a scalable architecture with a well-defined two-level system. The spin of an electron (or $I$=1/2 nuclear moment) forms a natural two-level system that cannot "leak" into other degrees of freedom. In recent years, there has been a great deal of interest in exploring and exploiting the spin degree of freedom in electronic systems. With conventional electronics, the spin of the electron is essentially irrelevant; the importance of the spin degree of freedom is epitomized in the way that the electron wavefunction is commonly expressed (see Eq. 1(a)). By contrast, with so-called spintronics, it is the spin degree of freedom that contains the relevant quantum information; the spatial part of the wavefunction provides a convenient "handle" by which spins are transported and made to interact (see Eq. 1(b)).

Eq. 1(a) $$\Psi_e = \psi(\vec{r}) \cdot \chi_s$$

Eq. 1(b) $$\Psi_e = \chi_s \cdot \psi(\vec{r})$$

**Requirement (R2)** is important for "booting" the quantum computer. The problem of initializing a collection of spins in semiconductors to a specific repeatable state is non-trivial. One way to achieve this result is to place them in a magnetic field at low temperature. To achieve sufficiently high polarization in bulk Si using a magnetic field strength $B$=1 Tesla, very low temperatures ($T$~100 mK) are required. Because the equilibrium spin polarization is a function of $B/T$, larger fields could be used at higher temperatures. However, smaller fields are preferable because many decoherence mechanisms scale as higher powers of $B$ [12]. Optical pumping of spin-polarized carriers is a relatively straightforward method for producing spin-polarized electrons in direct gap semiconductors like GaAs[13]. Unfortunately, all-optical approaches are not simply extended to indirect gap semiconductors such as Si and Ge. Direct optical transitions can be engineered in Si by creating small Ge/Si quantum dots; these quantum dots may provide a method for meeting requirement (R2).

**Requirement (R3)** states that there must be a separation of timescales (by at least $10^5$) that characterize the gate time and the decoherence time in the system. Both time scales depend on the physical system, as well as the way in which the subsequent requirement (R4) will be satisfied. The weak spin orbit coupling in Si ($|g-2|<3 \times 10^{-3}$) leads to reduced $T_2$ times on the order of $10^{-4}$-$10^{-5}$ s for donor electrons in Si[14]. Strain and the introduction of other materials such as Ge are expected to lead to reduced values for $T_2$; nevertheless, these times are still long compared to the best values reported for GaAs[4].

**Requirement (R4)** relates to the central processing unit of the quantum computer—the quantum logic gates that process quantum information. A minimal set of one-qubit and two-qubit gates is required for quantum computation.

There are two ways in which ferroelectric materials will be used in satisfying (R4). The first has to do with controlling the static interactions between electrons through ferroelectric domain patterning. Ferroelectric materials are useful "hands" that can manipulate the charge degree of freedom with the necessary high spatial and temporal resolution required for quantum information processing. Recent materials advances by McKee *et al.*[15] have resulted in high quality *crystalline* interfaces between oxide ferroelectrics and semiconductors. Bulk $BaTiO_3$ has a spontaneous polarization $P_s$=26 µC/cm$^2$ at room temperature, corresponding to a sheet density $\sigma=\pm1.6\times10^{14}$ cm$^{-2}$. The field effect at the ferroelectric/semiconductor interface is so strong that quantum confinement effects should be achievable with suitably small domain patterning. The second property of ferroelectric materials is their nonlinear response to electromagnetic fields. Optical rectification, a manifestation of the linear electrooptic effect, allows an optical field to be converted into a terahertz polarization that persists as long as the optical pulse is present. Terahertz ferroelectric switches may allow the necessary control over electron spin interactions required to create exchange gates.

Recent theoretical developments by DiVincenzo *et al.*[16] have shown that it is possible to use decoherence-free subspaces of multi-spin registers as composite qubits, and that effective one-qubit and two-qubit interactions can be implemented using only (Heisenberg) exchange gates. There are two clear advantages of this approach to quantum computation. First, there is significant overhead involved in developing a quantum gate, and an approach that simplifies the hardware implmentation is clearly welcome. Second, the exchange interaction is in principle very fast, if it can be controlled with sufficient precision.

**Requirement (R5)** delineates the requirements for final state measurement. Optical methods such as Faraday rotation have a low "quantum efficiency", and cannot measure single quantum events with high fidelity. It is most likely that electronic methods, perhaps involving single electron transistors, will provide the necessary means for satisfying this requirement. It could be argued that a quantum information processor need only satisfy requirements (R1-4), or even (R1,3-4).

## III. Material Systems

The viability of the proposed approach to quantum information processing depends critically on the material systems and their properties. Below I discuss two important subsystems and their relevant properties.

### A.   Ge/Si Quantum Dots

Bulk Si and Ge are indirect gap semiconductors that absorb or emit photons only with the assistance of a phonon to conserve momentum. As a consequence, the optical selection rules that enable spin-polarized carriers to be created in direct band semiconductors like GaAs are absent for such materials[17]. In order to initialize the state of a quantum computer (requirement (R2)), it will be important to engineer direct optical transitions in Si using Ge quantum dots. These quantum dots can also be used to localize the electrons.

One way to convert an indirect semiconductor into a direct one is to form a superlattice (*i.e.*, Si and Ge)[18]. In doing so, the size of the Brillouin zone is reduced and a portion of the conduction band minimum is folded onto the $\Gamma$ point, resulting in a quasidirect gap system. For semiconductor quantum dots, a momentum space representation is not appropariate, but if the envelope wavefunctions of confined carriers have a sizable Fourier component at the wavevector corresponding to the direct gap transition, then direct optical transitions become allowed. Takagahara and Takeda[19] have calculated radiative rates for Ge quantum dots, showing an indirect-direct crossover near a diameter $d$~10 nm. Naturally formed Si/Ge quantum dots grown by self-assembly using the Stranski-Krastanov mode have a range of diameters that start at approximately 20 nm and increase to as large as 50-100nm. These quantum dots are *not useful* for quantum computing applications, because they are too large to have appreciable direct optical transitions. In order to grow quantum dots with smaller sizes, several groups have deposited a nucleation layer, using either semiconducting (C)[20,21] or n-type (Sb)[22] materials; both have been demonstrated to yield quantum dots with strong luminescence and signatures of direct optical transitions[22,23]. It has also been shown[20] that conditions exist in which quantum dots will grow spontaneously only when Carbon is present. The small diameter and bright luminescence of these quantum dots are encouraging for optical control over the initial spin state of the electrons.

**B.    Ferroelectric Thin Films**

Below, I review some relevant aspects of nonlinear optics that are central to the proposed quantum information processor, and discuss signal strengths for reasonable parameter values.

*Optical Rectification*

The induced polarization $P$ in a nonlinear material is conventionally expanded in powers of the electric field strength $E$:

Eq. 2.  $P_i(t) = \chi^{(1)}_{ij} E_j(t) + \chi^{(2)}_{ijk} E_j(t) E_k(t) + ... \equiv P^{(1)}_i(t) + P^{(2)}_i(t) + ...$

The second order contribution leads to second harmonic generation (SHG), difference frequency generation (DFG), and optical rectification (OR). Using the complex representation $\widetilde{E}(t) = E_1 e^{-i\omega_1 t} + E_2 e^{-i\omega_2 t} + c.c.$, one obtains for the case of DFG:

Eq. 3 $$P_i^{(2)}(\omega_1 - \omega_2) = 2\chi_{ijk}^{(2)} E_{1j} E_{2k}^*.$$

For the case of a femtosecond pulse, the various Fourier contributions to DFG interfere to give a temporal profile for the nonlinear polarization that is proportional to the optical intensity $I_{opt}$:

Eq. 4 $$P^{(2)}(t) = \frac{rn^3}{2c} I_{opt}(t),$$

where $r$ is the electrooptic coefficient (indices suppressed for simplicity), $n$ is the refractive index, and $c$ is the speed of light. The numerical value of the maximum value for the polarization is expressed in terms of quantities relevant for $BaTiO_3$ and accessible in the laboratory:

Eq. 5. $$P_{max}^{(2)} = (6.29 \times 10^{-2}\,\mu C/cm^2) \left(\frac{I_{avg}}{10mW}\right) \left(\frac{D}{\mu m}\right)^2 \left(\frac{76MHz}{\Omega}\right) \left(\frac{r}{1.95 \times 10^{-11} m/V}\right) \left(\frac{\tau_{opt}}{100\,fs}\right) \left(\frac{n}{2.45}\right)^3$$

where $D$ is the laser spot diameter, $\Omega$ is the laser repetition rate, $r$ is the electrooptic coefficient, $\tau_{opt}$ is the pulse width, and $n$ is the refractive index. Because the nonlinear polarization arises from a non-resonant process below the optical gap of the semiconductor (and ferroelectric), it is possible to use large power densities without concern for photogeneration of carriers, heating, etc. For $BaTiO_3$, using a diffraction-limited spot with 10 mW average power, one obtains the following nonlinear polarization for $BaTiO_3$:

$$P_{max}^{(2)} = 6.29 \times 10^{-2}\,\mu C/cm^2 = 3.93 \times 10^{12}\,e^-/cm^2.$$ Because ferroelectric domains can be patterned

with high precision, and because the sign of the electrooptic coefficient depends on the polarization direction, *the spatial extent of the optically induced electric fields is not limited by diffraction*. That is, the nonlinear polarization may be used to actuate a gate separating two electrons using optical fields that extend over much larger length scales.

Generation of propagating terahertz electromagnetic waves using optical rectification was achieved by Nahata and Heinz, who bonded a $LiTaO_3$ superstrate to a coplanar microstripline[24]. Using modest energy densities, they measured significant voltages (>0.7 mV) 350 μm from the source. Pulse broadening and the effects of bonding both conspire to lower the effect. In epitaxial films, the separation between ferroelectric and semiconductor is measured in Angstroms—hence, the effects are expected to be significantly larger for epitaxial ferroelectrics on semiconductors.

*Local Magnetic Fields*

The transient displacement currents from ferroelectric domains will induce transient magnetic fields. While the exact profile depends on the details of the domain pattern, the overall scale can be estimated by considering a simple geometry, in which a circular cylinder of radius *R* sustains a displacement current density of magnitude $P^{(2)}_{max}/\tau_{opt}$, as shown in Figure 2. The maximum field strength is given by Eq. 6. For reasonable parameters, the peak magnetic field can be quite large; however, the temporal profile, which is proportional to the time derivative of the optical intensity $I_{opt}(t)$, limits its use for electron spin resonance. The combination of electric and magnetic fields may lead to useful geometric phases (*i.e.*, Berry's phase) that can be exploited in single-spin phase shifts; however, significantly more detailed modelling of the physical system is required to assess the strength and usefulness of these effects. As a source of decoherence (systematic phase errors), it will be important to understand these contributions.

Eq. 6
$$B_{max} = \frac{\mu_0 R P^{(2)}_{max}}{\tau_{opt}} = (39.6 \text{ gauss}) \left( \frac{R}{0.5 \text{ μm}} \right) \left( \frac{P^{(2)}_{max}}{6.29 \times 10^{-2} \text{ μC/cm}^2} \right) \left( \frac{100 \text{ fs}}{\tau_{opt}} \right)$$

In order to perform arbitrary quantum logic operations, a "universal" set of gates is required[25]. Until recently, most attention has been concerned with identifying mechanisms that will enable both single qubit rotations and nearest neighbor exchange gates. However, DiVincenzo *et al.* have recently shown that the exchange gate alone is universal when multi-spin registers form the fundamental qubit[16]. Using spatiotemporal pulse shaping methods, it should be possible to control the spacing and timing of these terahertz pulses and create the fundamental gate operations required for quantum information processing.

## IV. Quantum Logic Operations

### A. The quantum swap

At the heart of the quantum information processor is a mechanism for controlling spin exchange between neighboring electrons. This interaction is achieved through the combination of static and terahertz fields of an epitaxial ferroelectric thin film. Ferroelectrically defined nanowires ensure that the electrons reach the tunneling barrier without "wandering". Figure 3 depicts the transient potential induced by optical rectification, and the induced exchange interaction through a ferroelectrically defined tunneling barrier. The top portion of each frame in Figure 3 depicts the one-dimensional channel along which the electrons travel and interact. At the center, the channel width is narrower, giving rise to a tunneling barrier. The barrier can arise from both quantum confinement along the transverse direction and from the electrostatically defined linewidth. The bottom portion shows the electrostatic potential along the center of the nanowire. Two potential wells each hold a single electron. Also shown are the two lowest energy solutions to the one-dimensional Schrödinger equation. Within the single-electron approximation, these solutions indicate the probability distribution for the electrons and their response to the transient coupling potential. The electrons are brought together using transient fields from a pair of femtosecond pulses centered on either side of the tunneling barrier. As these fields decay, a weaker third pulse is applied to the center barrier region, ensuring that the electrons return to their bound states.

The exchange operator $\hat{U}_{ex}(\theta) = \exp[-i\theta \hat{\mathbf{S}}_1 \cdot \hat{\mathbf{S}}_2]$ is characterized by an angle $\theta$ (see[8]) such that $\theta = \pi$ corresponds to a "swap" operation. In order to implement exchange-only universal gates[26], it is necessary to vary $\theta$ continuously from 0 to $2\pi$. Such control can be achieved by varying both the strength and duration of the optical pulses. Long-range coupling is provided through the use of sequentially applied swap operations.

In constructing an exchange gate such as the one in Figure 3, it is important to consider possible decoherence mechanisms. For instance, if the adiabatic condition is relaxed, the loss of spatial coherence can lead to decoherence of the spin wavefunction. One method of reducing such effects is to couple the spins *via* a superexchange mechanism[27]. By forcing electrons to communicate through a single quantum channel, such effects can be greatly reduced.

B. **Quantum Information Processing**

Once the basic one-qubit and two-qubit operations have been established, it will be important to "program" the quantum information processor to perform complex combinations of such operations. In recent years, Keith Nelson and coworkers at MIT have developed the optical technology for converting a femtosecond seed pulse into a controlled pattern of pulses shaped in both space and time[24]. At the core of this technology is a commercially available two-dimensional spatial light modulator (Hamamatsu PAL-SLM) that can be programmed using the standard VGA output of a PC video card. The algorithms for producing controlled waveforms of a desired pattern have also been developed[28]. At present, the speed with which the array could be reprogrammed will limited the interactive use of the SLM in a quantum calculation (the speed is quite slow, ~10 ms); however, if a repetitive calculation is performed, then it will not be necessary to reprogram continually the SLM.

## V. Summary

I have described a method for controlling electron spin interactions in Silicon-based structures using the static and dynamic polarization of an epitaxial ferroelectric thin film. Solid state approaches to quantum information processing and quantum computing have many advantages. In particular, Silicon-based approaches are especially attractive, given the natural interface to classical computing architectures. The question of whether it will be possible to satisfy the physical requirements for quantum computation can only be answered through experimental investigation.

## VI. Acknowledgements

# VII. References

[1] C. H. Bennett and D. P. DiVincenzo, Nature **404**, 247 (2000).

[2] C. H. Bennett and P. W. Shor, IEEE Trans. Inf. Theory (USA) **44**, 2724 (1998); C. H. Bennett, G. Brassard, S. Breidbart, et al., IBM Tech. Discl. Bull. (USA) **26**, 4363 (1984).

[3] P. D. Townsend, Nature **385**, 47 (1997).

[4] J. M. Kikkawa, I. P. Smorchkova, N. Samarth, et al., Science **277**, 1284 (1997).

[5] J. M. Kikkawa and D. D. Awschalom, Nature **397**, 139 (1999).

[6] I. Malajovich, J. M. Kikkawa, D. D. Awschalom, et al., Phys. Rev. Lett. **84**, 1015 (2000).

[7] B. E. Kane, Nature **393**, 133 (1998).

[8] D. Loss and D. P. DiVincenzo, Phys. Rev. A **57**, 120 (1998).

[9] R. Vrijen, E. Yablonovitch, K. Wang, et al., quant-ph/9905096 (1999).

[10] D. P. Divincenzo, quant-ph/0002077 (2000).

[11] J. Preskill, http://www.theory.caltech.edu/~preskill/ph229 (1999).

[12] A. Honig and E. Stupp, Phys. Rev. **117**, 69 (1960); D. K. Wilson, Phys. Rev. **124**, 1068 (1961).

[13] D. Deutsch, Proc. R. Soc. Lond. A, Math. Phys. Sci. (UK) **400**, 97 (1985).

[14] J. P. Gordon and K. D. Bowers, Phys. Rev. Lett. **1**, 368 (1958).

[15] R. A. McKee, F. J. Walker, and M. F. Chisholm, Phys. Rev. Lett. **81**, 3014 (1998).

[16] D. P. DiVincenzo, D. Bacon, J. Kempe, et al., quant-ph/ 0005116 (2000).

[17]B. Lax and Y. Nishina, Phys. Rev. Lett. **6**, 464 (1961).

[18]T. P. Pearsall, J. Bevk, L. C. Feldman, et al., Phys. Rev. Lett. **58**, 729 (1987).

[19]T. Takagahara and K. Takeda, Phys. Rev. B **46**, 15578 (1992).

[20]O. G. Schmidt, A. C. Lange, K. Eberl, et al., Appl. Phys. Lett. **71**, 2340 (1997).

[21]S. Schieker, O. G. Schmidt, K. Eberl, et al., Appl. Phys. Lett. **72**, 3344 (1998).

[22]D. Gruetzmacher, R. Hartmann, O. Leifeld, et al., in *Silicon-based Optoelectronics* (Int. Soc. Opt. Eng., San Jose, CA, 1999), p. 171.

[23]C. S. Peng, Q. Huang, W. Q. Cheng, et al., Phys. Rev. B **57**, 8805 (1998).

[24]R. M. Koehl, T. Hattori, and K. A. Nelson, Optics Communications **157**, 57 (1998).

[25]A. Barenco, C. H. Bennett, R. Cleve, et al., Phys. Rev. A **52**, 3457 (1995).

[26]A. Nahata and T. F. Heinz, Optics Letters **23**, 867 (1998).

[27]P. Recher, D. Loss, and J. Levy, in *Macroscopic Quantum Coherence and Computing*, edited by D. Averin and P. Silverstrini (Plenum Press, New York, Naples, Italy, 2000) (cond-matt/0009270).

[28]Keith Nelson, private communication.

FIGURE CAPTIONS

Figure 1. Schematic of ferroelectric domains and optical rectification. (a) $I_{opt}$=0. (b) $I_{opt}$>0. The magnitude of the ferroelectric polarization decreases upon illumination.

Figure 2. (a) Geometry for estimating the magnetic field strength due to ferroelectric displacement currents. (b) Schematic of temporal profile for magnetic field.

Figure 3. Quantum swap operation. Electrons are guided along a ferroelectrically defined nanowire (top), with a tunneling "kink" at the center. Bottom graphs show 1-d effective potential along the nanowire. Optical rectification of a femtosecond pulse generates a transient coupling between quantum dots. (a) t= -200 fs. (b) t= -100 fs. (c) t=0. Ionized electrons undergo controlled exchange. (d) t= 150 fs. Coupling is turned off. (e) t= 450 fs. Weaker second pulse centered on tunneling barrier returns electrons to their confined states. (f) t= 600 fs. Controlled exchange operation is complete.

TABLE CAPTIONS

Table 1. Left, DiVincenzo's summary of the requirements for quantum computing [16]. Right, proposed method for meeting these requirements.

|      | **Requirement** | **Implementation** |
|------|-----------------|--------------------|
| (R1) | A scalable physical system with well-characterized qubits | Electron spins in Si |
| (R2) | The ability to initialize the state of the qubits to a simple fiducial state, such as \|000…0> | Optical spin injection into Si using quasi-direct gap Ge quantum dots |
| (R3) | Long relevant decoherence times, much longer than the gate operation time | Long spin lifetimes in Si; fast (2-qubit) gate operation times using ferroelectric gates |
| (R4) | A "universal" set of quantum gates | One and two-qubit operations possible |
| (R5) | The ability to measure specific qubits | Single electron transistors (non-optical) |

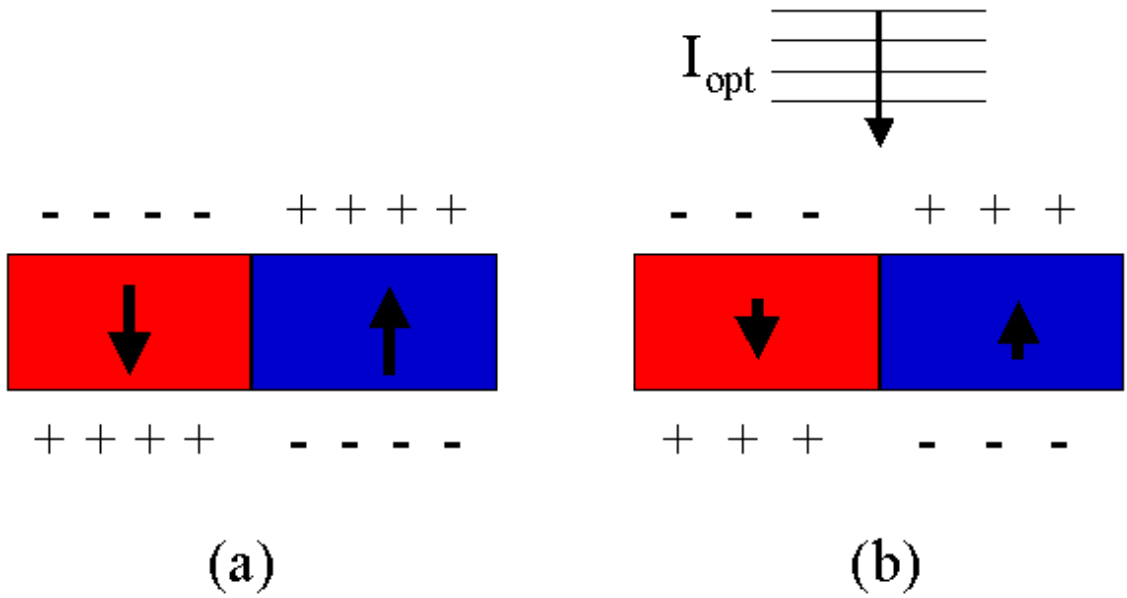

Levy, "Quantum Information Processing…", Figure 1

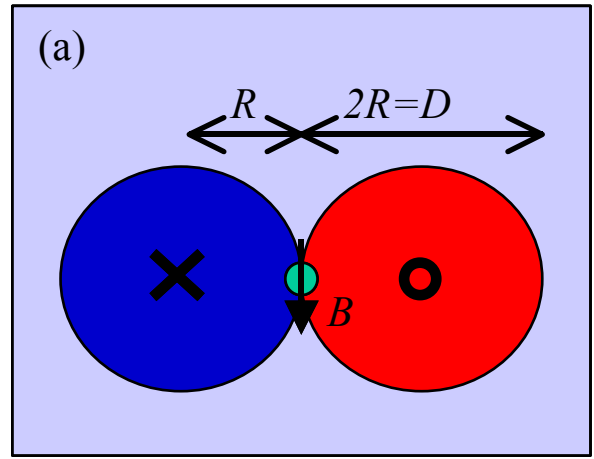

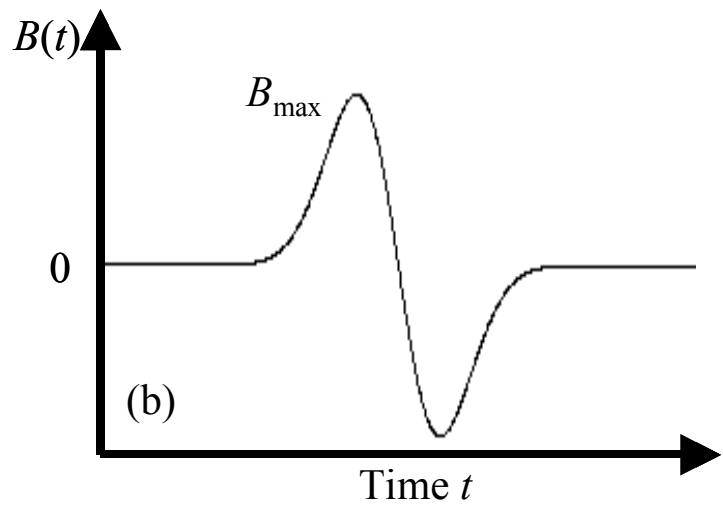

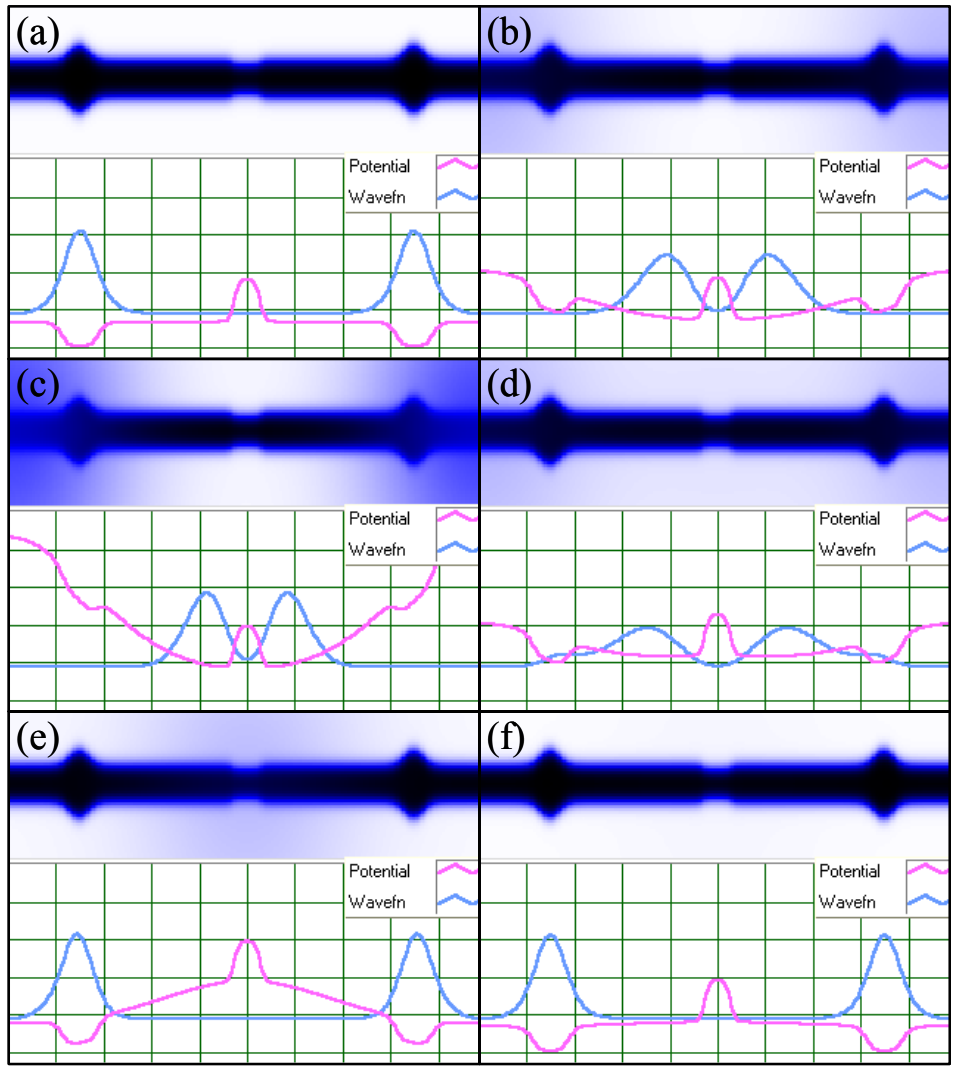